\documentclass[journal]{IEEEtran}
\usepackage{graphicx}

\begin{document}
\thispagestyle{empty}
%

\title{Artificial Neural Network as a FPGA Trigger \\for a Detection of Very Inclined Air Showers}
\author{Zbigniew~Szadkowski,~\IEEEmembership{Member,~IEEE}, K. Pytel %
\thanks{Manuscript received June 27, 2014.  
This work was supported by the Polish National Center for
Research and Development under NCBiR Grant No. ERA/NET/ASPERA/02/11
and by the National Science Centre (Poland) under NCN Grant No. 2013/08/M/ST9/00322
}
\thanks{Zbigniew~Szadkowski is with the University of \L{}\'od\'{z}, 
Department of Physics and Applied Informatics, 
Faculty of High-Energy Astrophysics, 90-236 \L{}\'od\'{z}, 
Pomorska 149, Poland, 
(e-mail: zszadkow @kfd2.phys.uni.lodz.pl, phone: +48 42 635 56 59).}
\thanks{Krzysztof Pytel is with the University of \L{}\'od\'{z}, 
Department of Physics and Applied Informatics, 
Faculty of Informatics, 90-236 \L{}\'od\'{z}, Poland,}%
}
\maketitle

\begin{abstract}

The observation of ultra-high energy neutrinos (UHE$\nu$s) 
has become a priority in experimental astroparticle physics. 
Neutrinos can interact in the atmosphere 
(downward-going $\nu$) or in the
Earth crust (Earth-skimming $\mathrm{\nu}$), producing 
air showers that can be observed with arrays of detectors 
at the ground. The surface detector array of the Pierre Auger 
Observatory can detect these types of cascades. The 
distinguishing signature for
neutrino events is the presence of very inclined showers 
produced close to the ground (i.e., after having traversed 
a large amount of atmosphere).  Up to now, the Pierre Auger 
Observatory did not find any candidate for a neutrino event.
This imposes competitive limits to the diffuse flux of 
UHE$\mathrm{\nu}$s. 

A very low rate of events potentially generated by neutrinos 
is a significant challenge for a detection technique
and requires both sophisticated algorithms and high-resolution 
hardware. We present a trigger based on a pipeline artificial 
neural network implemented in a large FPGA which after 
learning can recognize traces corresponding to special 
types of events. 

The structure of an artificial neural network algorithm being 
developed on the MATLAB platform has been implemented 
into the fast logic of the biggest 
Cyclone\textsuperscript{\textregistered} V E FPGA used for 
the prototype of the Front-End Board for the Auger-Beyond-2015 
effort. Several algorithms were tested, however, the 
Levenberg-Marquardt one (trainlm) seems to be the most efficient. 

The network was taught: a) to recognize "old" showers 
(learning on a basis of real very inclined Auger showers 
(positive markers) and real standard showers especially triggered 
by Time over Threshold (negative marker), b) to recognize "young" 
showers (on the basis of simulated "young" events (positive markers) 
and standard Auger events as a negative reference).
A three-layer neural network being taught by real very inclined 
Auger showers shows a good efficiency 
in pattern recognition of 16-point traces with profiles 
characteristic of "old" showers. 

Nevertheless, preliminary simulations of showers with the CORSIKA 
shower simulation package and the response of the water Cherenkov 
tanks with the OffLine data analyses and reconstruction package
suggest that for neutrino showers starting a development deep 
in the atmosphere,
and for relatively low initial energy 
$\mathrm{\sim}$10$\mathrm{^{18}}$ eV, 
ADC traces are not too long,
and a 16-point analysis should be sufficient for a recognition 
of "young" showers.
The neural network algorithm can significantly support a detection 
for low energies, where a more intense neutrino stream is expected.
For higher energies traces are longer, however, a detector response 
is strong enough for the showers to be detected by standard 
amplitude-based triggers.

\end{abstract}


%
\IEEEpeerreviewmaketitle

\section{Introduction}

\IEEEPARstart{U}{ltra} 
-high energy cosmic ray (UHECR) experiments in energy range 
of $10^{18}-10^{20}$ eV are the inspiration for active development
of theoretical astrophysics hypotheses \cite{Nagano}. 
 The origin of the UHECRs, their production mechanism and 
composition still remain a mystery, as do hypothetical
fluxes of ultrahigh energy neutrinos (UHE$\nu$s) \cite{Halzen}.

Generally, we can classify astrophysical models as:  
''bottom-up''  and ''top-down''.
In the first case, protons and nuclei are accelerated in 
astrophysical shocks, while pions are
produced by cosmic ray interactions with matter or radiation 
at the source \cite{Becker}. 
The second scenario postulates that protons and neutrons are 
produced from quark and gluon fragmentation
of very heavy particles, according to Grand Unified Theories 
or Super-symmetries,
with a supremacy of pions over nucleons \cite{Sigl}. 
However, "top-down" models have been rather disfavored 
by the Pierre Auger Observatory 
due to relatively low photon limits \cite{photon_limit}.
Protons and nuclei also produce pions due to the 
Greisen-Zatsepin-Kuzmin (GZK) cutoff \cite{GZK}\cite{ZK}
seen by HiRes \cite{Abbasi} and confirmed by the 
Pierre Auger Observatory \cite{PAO} \cite{PAO-GZK}.

For primary protons, decays of charged pions (as results of photo-pion production associated with the GZK effect)
are expected to be the source of ultra-high energy neutrinos 
(UHE$\nu$s). However, their fluxes are still doubtful.
 If the primaries are heavy nuclei, the UHE$\nu$s should be 
significantly suppressed \cite{Olinto}.

The observation of UHE$\nu$s should support an explanation of the 
origin of UHECRs \cite{Seckel}. 
Neutrinos indicate directly the source of their production 
due to the absence of any deflection in magnetic fields.
Unlike photons, their unimpeded travel from the sources to the 
Earth may support a confirmation or rejection of production models.
UHE$\nu$s can be detected with arrays of detectors at ground 
level that are currently being used to measure 
extensive showers produced by cosmic rays, 
e.g. the Pierre Auger Observatory \cite{Zas}. 
The main challenge in this technique is the extraction of showers 
initiated by neutrinos from the "background" induced by 
regular cosmic rays.

Neutrinos have very small cross-sections. This implies a very 
low probability of interaction for any relatively thin target 
such as at small zenith angles of neutrino incidence, 
corresponding to a slant depth at the level 
of 1 $\mathrm{kg/cm^2}$. Neutrinos can interact at any point 
along their trajectories.
For a very inclined shower the slant depth increases to 
$\mathrm{\sim}$31 $\mathrm{kg/cm^2}$ and neutrino interactions 
become more probable \cite{Smirnov}.
Protons, nuclei, or photons usually interact shortly 
after entering the atmosphere. 
For inclined showers they produce a narrow muonic pancake 
(only the muonic component survives), while for
deeply interacting neutrinos the inclined showers contain 
also a significant electromagnetic contribution.
Inclined showers that interact deep in the atmosphere may 
be a signature of neutrino events.
The surface detector array (SD) of the Pierre Auger Observatory 
can detect both Earth-skimming and down-going showers \cite{AHEP}.
The Earth skimming neutrino events are limited to a very 
narrow zenith range where the expected  background 
of nucleonic showers is very small. For downward-going neutrino 
showers there is in principle a larger range of possible 
zenith angles, but with larger background contamination. 
This imposes specific algorithms allowing a separation of 
neutrino-induced showers from nucleonic ones.

\begin{figure}[!t]
\centering
\includegraphics[width=\columnwidth,keepaspectratio]
{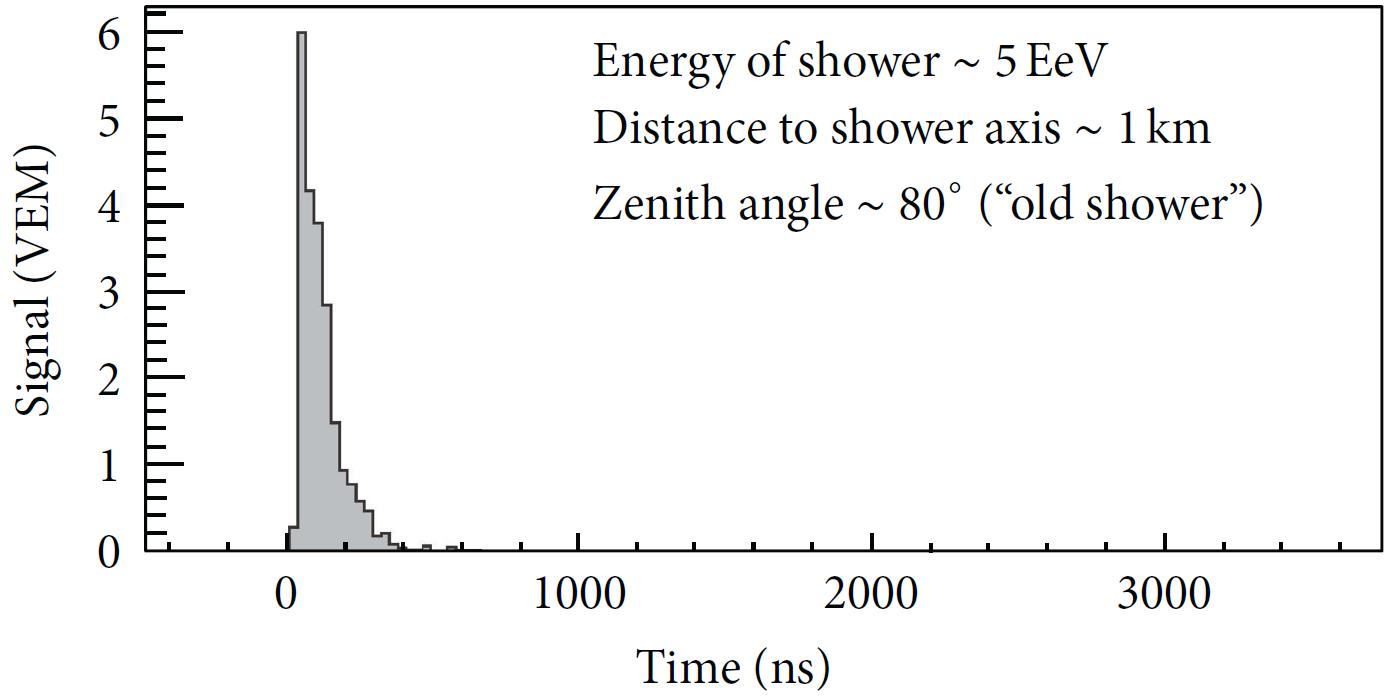}
\includegraphics[width=\columnwidth,keepaspectratio]
{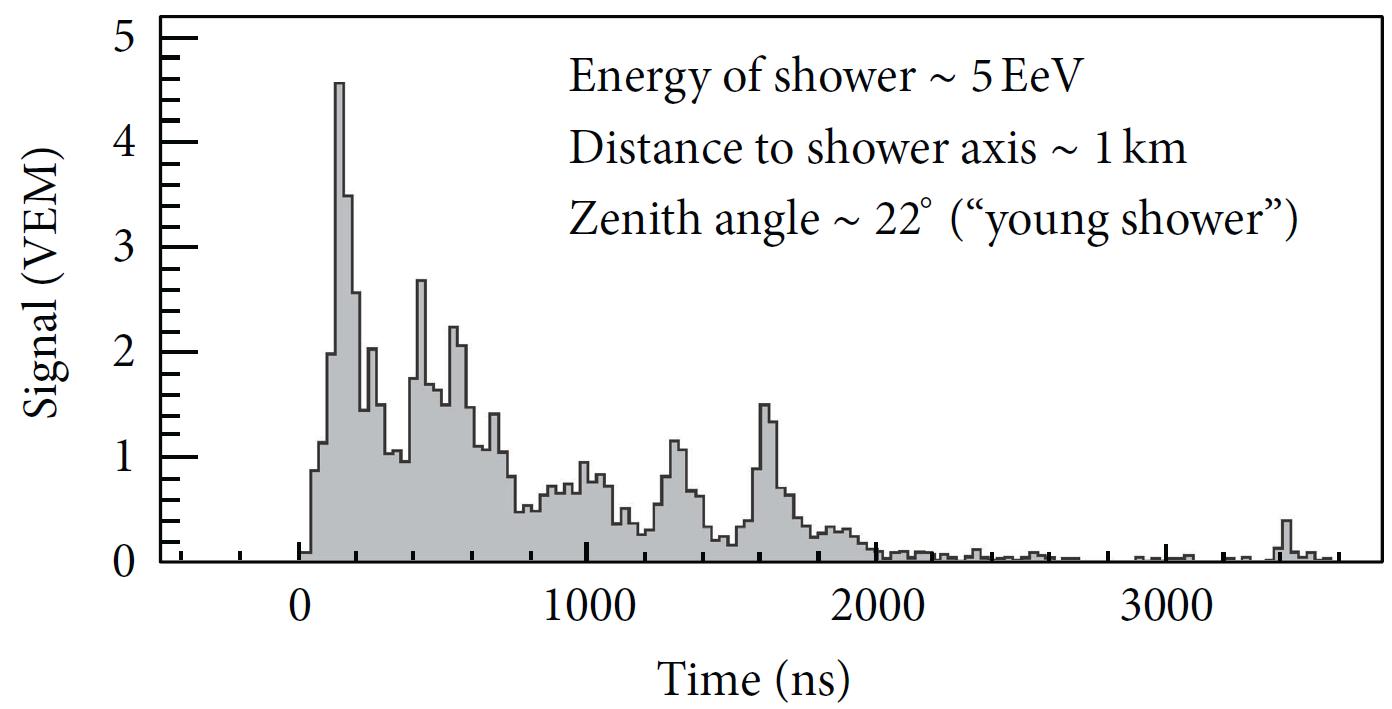}
\caption{Simulated ADC traces of stations at 1 km from 
the shower core for two real showers of 5 EeV. 
(a) “Old” extensive air shower ($\theta \sim ∼ 80^{\circ}$). 
(b) Shower arriving in the early stages of
development (“young” shower) \cite{AHEP}. }
\label{sim_traces}
\end{figure}

In the downward-going channel neutrinos can be generated 
via both charged and neutral current interactions. 
Neutrinos of any flavor can induce
extensive air showers along the entire path of their 
development in the atmosphere, also
very close to the ground \cite{Capelle}.

In the Earth-skimming channel showers can be induced by 
$\nu_{\tau}$ being a product of a $\tau$
lepton decaying after the propagation and interaction of an
upward-going $\nu_{\tau}$ inside the Earth \cite{Bertou}. 

The surface detector of the Pierre Auger Observatory has good 
potential to identify and 
to separate neutrino-induced showers (for both the Earth-skimming 
and downward-going channels) 
from showers induced by regular cosmic rays for a large zenith angle ($\mathrm{\theta \ge 70^{\circ}}$).
One of the fundamental criteria allowing an identification 
of neutrino-induced showers is 
the timing of shower fronts directly observed as profiles 
of registered traces in the surface detectors.

\section{ADC traces analysis}

The surface detector array of the Pierre Auger Observatory 
is able to detect and identify UHE$\nu$s for 
E $\ge$ 10$^{18}$ eV \cite{tau}. 
Due to a much larger first interaction cross-section for 
protons, heavier nuclei and even photons than for neutrinos, 
their showers usually appear shortly after
entering the atmosphere. However, neutrinos can generate 
showers deeply into the atmosphere. 
Vertical showers initiated by protons or heavy nuclei have 
a considerable amount of electromagnetic component at
the ground ("young" shower front). However, at high zenith 
angles ($\mathrm{\theta \ge 70^{\circ}}$)
(thicker than about three vertical atmospheres), UHECRs 
interacting high in the atmosphere generate shower
fronts dominated by muons at the ground (an "old" shower front), 
which generate narrow signals (short ADC traces) 
spreading over typically tens of nano-seconds in practically 
all the stations of the event.
These traces can be recognized with an algorithm of a 16-point 
discrete cosine transform (DCT)  
as well as with a 16-point input artificial neural network (ANN).

\begin{figure}[!b]
\includegraphics[width=0.9\columnwidth,keepaspectratio]
{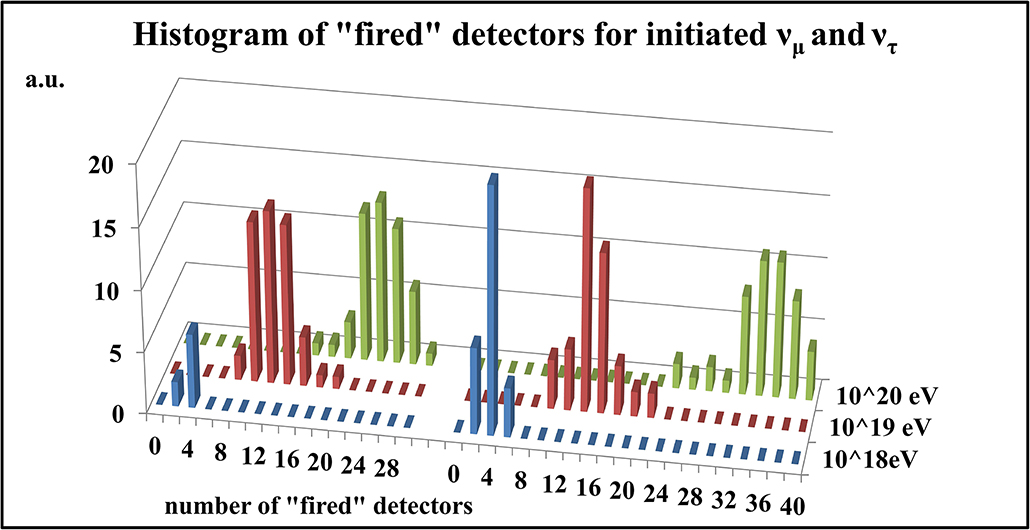}
\caption{ Histogram of "fired" surface detectors for showers 
initiated by $\nu_{\mu}$ (left panel)and $\nu_{\tau}$ 
(right panel) neutrinos at an altitude of 9350 m.}
\label{mult}
\end{figure}

For a recognition of the very inclined "old" showers the 
DCT algorithm  \cite{DCT-NIM} was developed and tested 
on the SD test detector in Malarg\"ue (Argentina) 
\cite{CycloneIII} \cite{TNS2013-DCT}. The algorithm 
precisely recognized ADC traces of required shapes.
Up to now it was tuned for "old" showers, however, 
it could be optimized also for shapes 
characteristic for "young" showers.
The ANN algorithm is an alternative approach. 
The efficiency of both algorithms will be tested
for both types of showers.

"Young" showers are spread in time over thousands of 
nano-seconds (Fig. \ref{sim_traces}) \cite{AHEP}.
For the "old" showers practically only the muonic 
component survives. It gives a short bump in the SD.
The "young" showers comprise also some electromagnetic 
component, which spread the ADC traces in time. 
However, the muonic component of "young" showers is 
ahead of the electromagnetic one and gives an early bump.
The rising edge of the bump is not so sharp as for the 
"old" ones, but this signal is also relatively rapidly 
attenuated, until the electromagnetic component starts 
to provide its own contribution. 
The ANN approach can focus on the early bump,
to select traces potentially generated by neutrinos. 

\begin{figure}[!t]
\centering
\includegraphics[width=\columnwidth,keepaspectratio]
{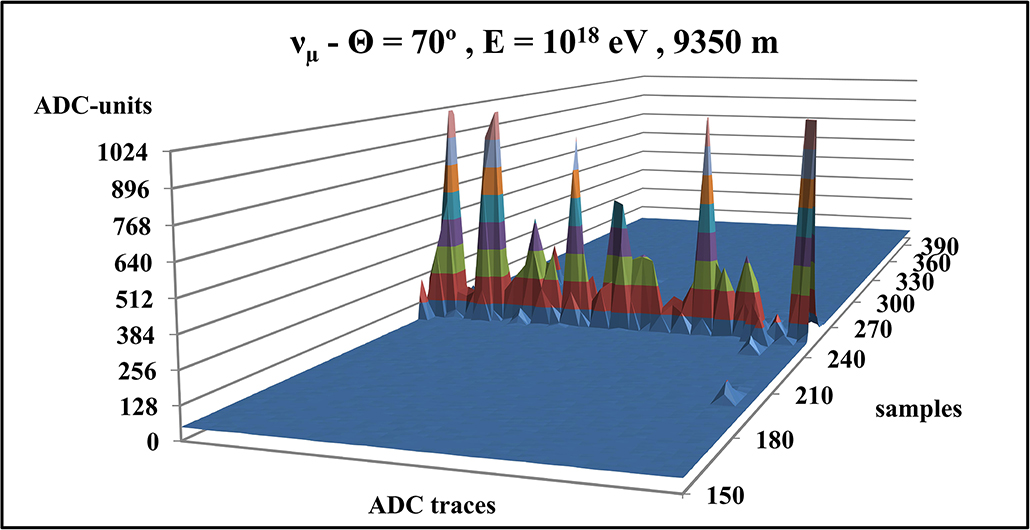}
\includegraphics[width=\columnwidth,keepaspectratio]
{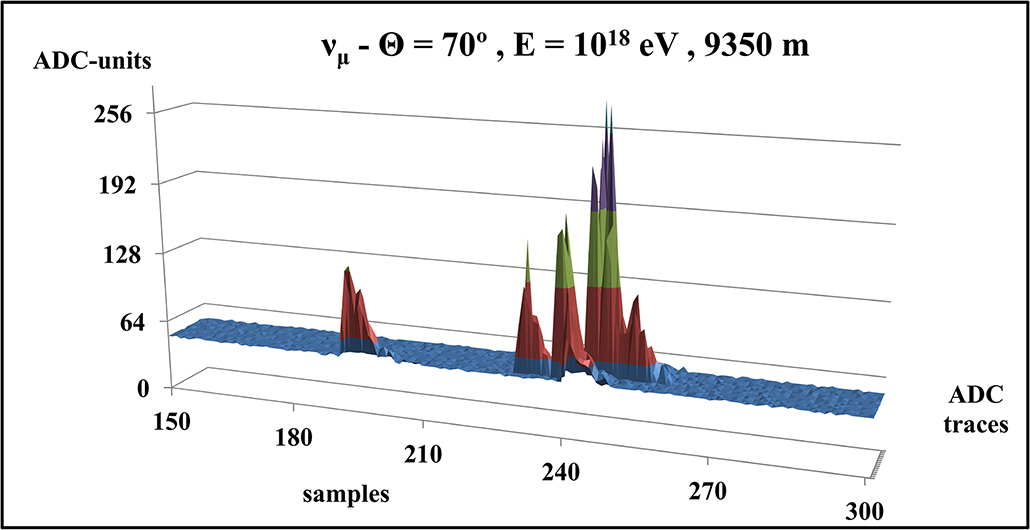}
\includegraphics[width=\columnwidth,keepaspectratio]
{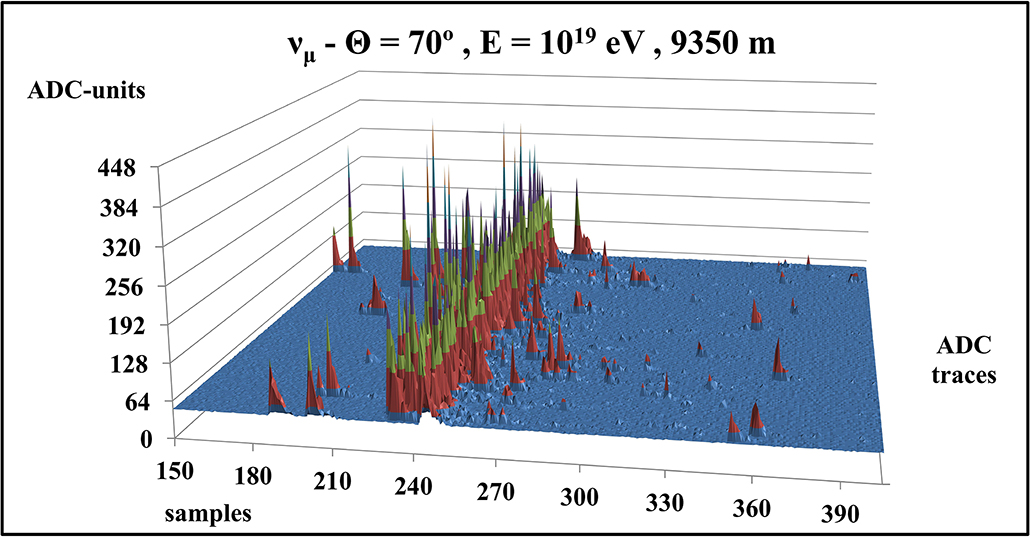}
\includegraphics[width=\columnwidth,keepaspectratio]
{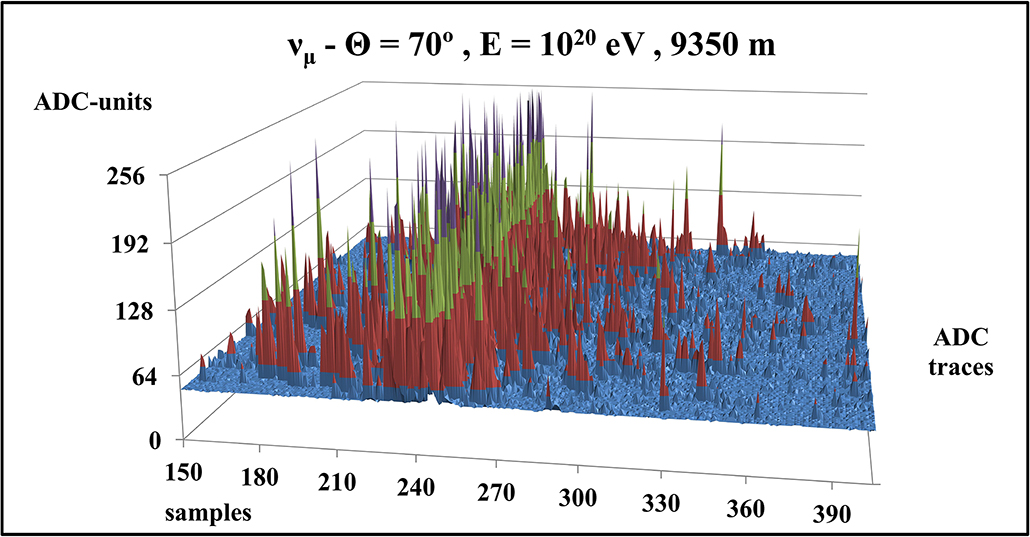}
\caption{Simulated ADC traces for $10^{18}$,  
$10^{19}$ and $10^{20}$ eV, respectively, 
for an incident $\nu_{\mu}$ at 9350 m
and  $70^{\circ}$ zenith angle. Shown here are events 
not giving strong enough signals to be detected 
by a standard trigger.}
\label{muon-sim}
\end{figure}

\begin{figure}[!t]
\centering
\includegraphics[width=\columnwidth,keepaspectratio]
{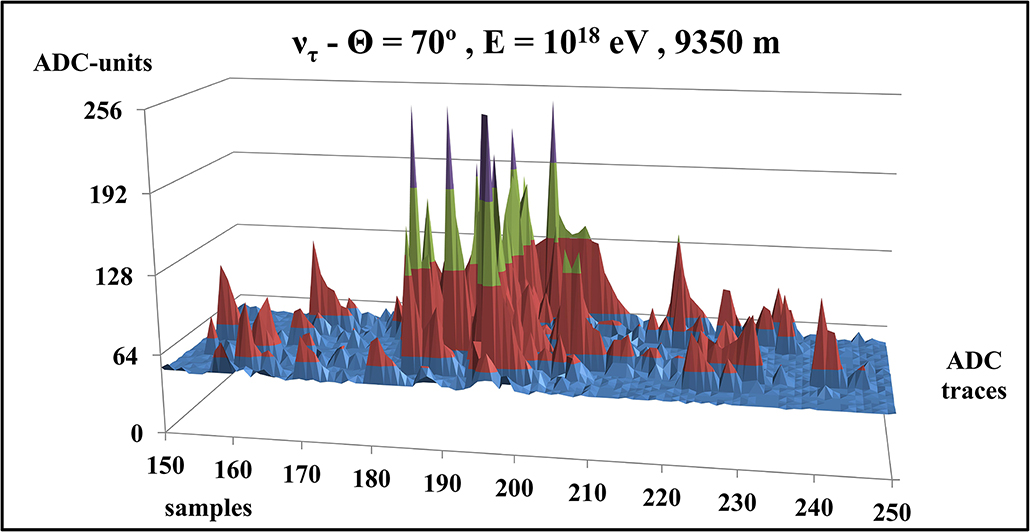}
\includegraphics[width=\columnwidth,keepaspectratio]
{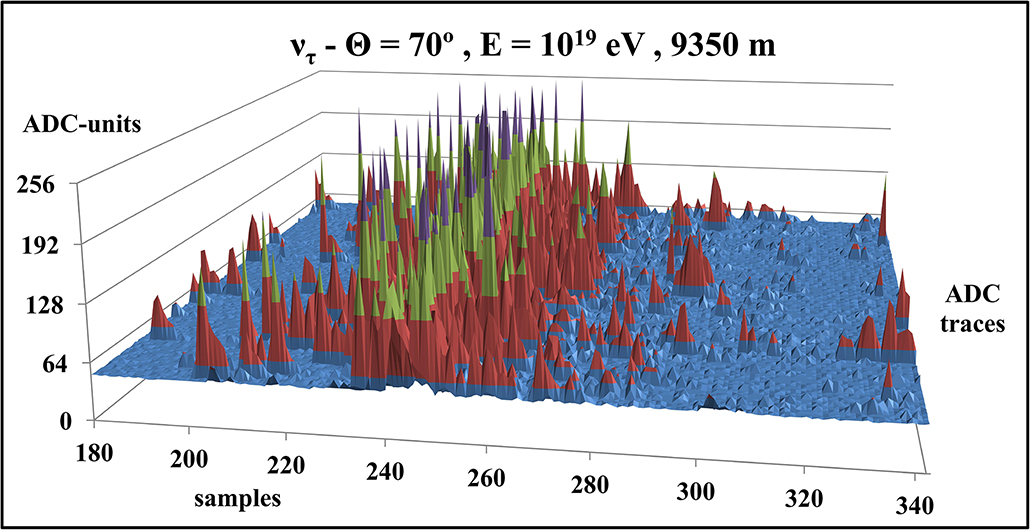}
\includegraphics[width=\columnwidth,keepaspectratio]
{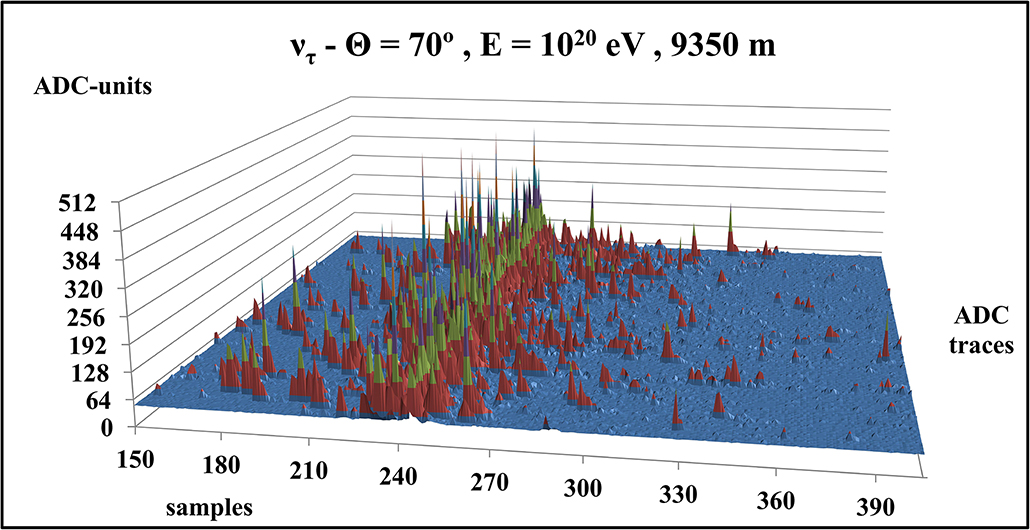}
\caption{Simulated ADC traces for $10^{18}$,  
$10^{19}$ and $10^{20}$ eV, respectively, for an 
incident $\nu_{\tau}$ at 9350 m
and  $70^{\circ}$ zenith angle. Shown are events with 
signals too low to be detected by a standard trigger.}
\label{tau-sim}
\end{figure}

\begin{figure}[!b]
\centering
\includegraphics[width=\columnwidth,keepaspectratio]
{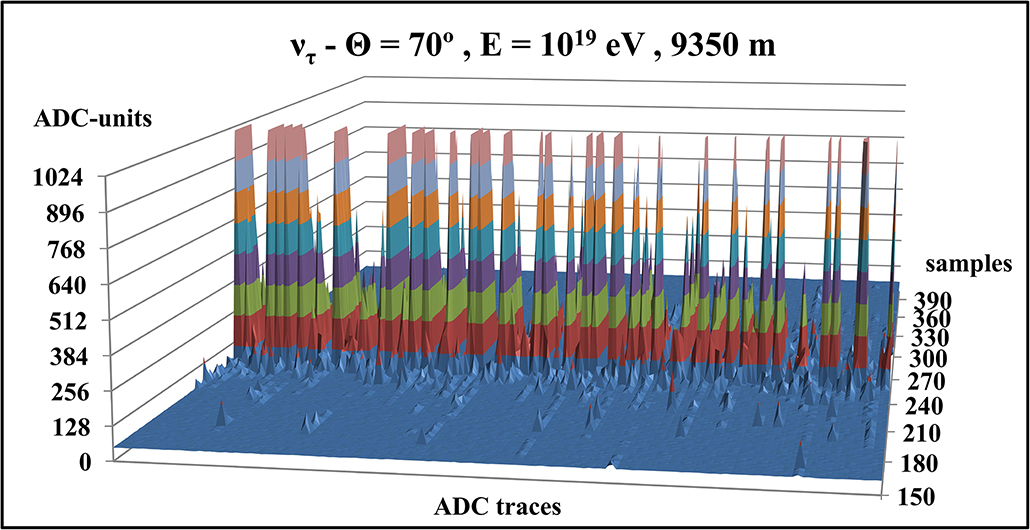}
\caption{Simulated ADC traces for incident $\nu_{\tau}$, 
for energy =  $10^{19}$ eV at 9350 m and  
70$^{\circ}$ zenith angle. }
\label{saturation}
\end{figure}

\begin{table}[t]
\centering\caption{\label{3-fold} Percentage of simulated 
events with traces detected by 3-fold coincidences 
(a standard threshold trigger in a time domain)}

\begin{tabular}{|c|c|c|c|c|}
\hline
 Energy& $\mathrm{\nu_{\mu}}$         &    $\nu_{\tau}$        &       $\nu_{\mu}$         &   $\nu_{\tau}$         \\
               &$\mathrm{\theta=22^{\circ}}$&$\mathrm{\theta=22^{\circ}}$&$\mathrm{\theta=70^{\circ}}$&$\mathrm{\theta=70^{\circ}}$ \\
\hline \hline                                                   
$\mathrm{10^{18}\;eV}$  &  33 \%  &  10 \% &  83 \%  &  75 \%                \\
$\mathrm{10^{19}\;eV}$  &  53 \%  &  40 \% &  89 \%  &  88 \%                 \\
$\mathrm{10^{20}\;eV}$  &  80 \% &  83 \% &  87 \%  &  88 \%              \\
\hline                                                   
\end{tabular}
\end{table}

On the other hand, independent simulations of showers in 
CORSIKA \cite{CORSIKA} 
and a calculation of a response of water Cherenkov 
detectors (WCDs) 
in OffLine \cite{OffLine} showed that for neutrino showers 
(initiated either by $\nu_{\mu}$ or $\nu_{\tau}$) for 
relatively large zenith angle (e.g. $70^{\circ}$) and 
low altitude (9 km) (to be treated as "young" showers 
before the maximum of development) there are relatively 
short ADC traces which can be
analyzed also by 16-point pattern engines.

Showers induced by relatively low-energy neutrinos 
(in the range of $10^{18}$ eV) (independently of flavor) 
can "fire" only a few surface detectors (Fig. \ref{mult}). 
These showers may be ignored due to the Auger T3 trigger 
\cite{trigger},
although they can generate even saturated traces in 
a few surface detectors (Fig. \ref{saturation}).

Table \ref{3-fold} shows a fractional rate of simulated events 
giving 3-fold coincidences on the T1 threshold trigger \cite{trigger}.
For low energy neutrino showers a parallel trigger based on a pattern recognition (i.e. using an artificial neural network)
can improve the probability of neutrino-induced shower detection.

Figs. \ref{muon-sim} and \ref{tau-sim} show that ignored events 
(which do not obey a condition of the T1 threshold trigger),
especially for low energies, can be analyzed by the 16-point 
only algorithm, either the DCT or the ANN one.

\begin{figure}[b]
\centering
\includegraphics[width=\columnwidth,keepaspectratio]{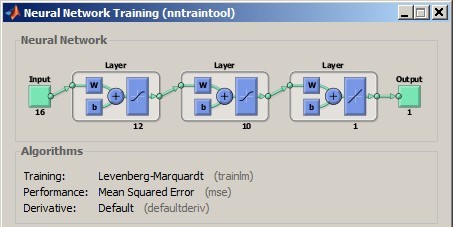}
\caption{An example of the ANN structure used for an 
optimization in Neutral Network Training 
of the MATLAB toolbox. }
\label{NNT}
\end{figure}

\section{MATLAB analysis}

The main motivation of an ANN implementation as a shower 
trigger is the fact that up to now the entire array did not 
register any neutrino-induced event.The probable reasons 
are: a) a very low flux of neutrinos 
and b) amplitudes of ADC-traces that are small and probably 
below the threshold of the standard 3-fold coincidence trigger.
The main idea is to use the ANN approach as 
a pattern recognition technique. 

Several networks were tested (Fig. \ref{NNT}) to get 
a reasonable compromise between the efficiency 
of the pattern recognition and a resource occupancy in the FPGA. 
To train the network we created a database 
of real Auger inclined "old" showers (as positive marker) 
and ''typical expected neutrino-like signal''
 (mostly vertical as negative markers).
Table \ref{methods} shows results for various teaching 
configurations used for two networks. Only the Levenberg-Marquardt
(Trainlm in MATLAB) algorithm was very efficient. The others showed unacceptable levels of error rates (Table \ref{methods}).

\begin{figure}[!t]
\centering
\includegraphics[width=\columnwidth,keepaspectratio]
{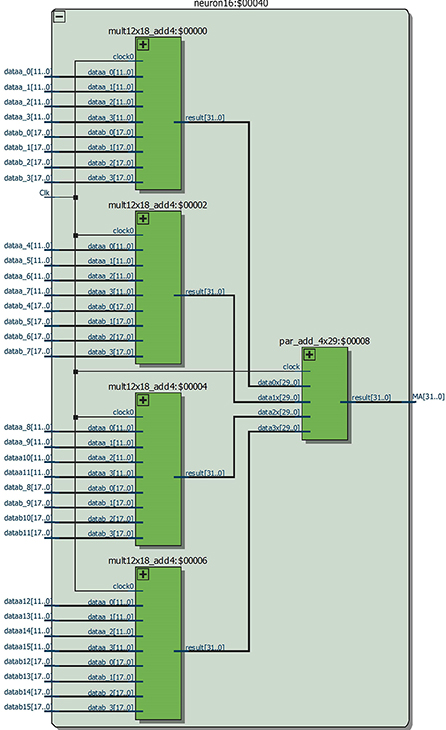}
\caption{An internal structure of an FPGA neuron.}
\label{neuron}
\end{figure}

Theoretically, a more complicated network could provide 
a higher efficiency (Table \ref{nets}), however, it requires 
much more FPGA resources, especially DSP embedded multipliers. 
The biggest FPGA from the 
Cyclone\textsuperscript{\textregistered} V E family - 5CEFA9F31I7 
contains 342 fast DSP embedded multipliers. For 3 independent 
ANN (for 3 PMTs in the Auger surface detector)
we can use 114 DSP blocks per channel. A single neuron 
(with sixteen 14-bit inputs and 14-bit coefficients) 
implemented with Altera\textsuperscript{\textregistered} 
Multiply Adder v13.1 and  
PARALLEL\_ADD Megafunctions requires 8 DSP blocks 
(Fig. \ref{neuron}). 114 DSP blocks allow an implementation 
of 14 neurons per PMT.

\begin{table}[b]
\centering\caption{\label{methods} Error rates for various training methods}

\begin{tabular}{|c|c|c|c|c|c|}
\hline
 ANN& cfg &Traincgp&Traingd&Trainlm &Trainscg  \\
\hline \hline                                                   
                  &Total    &  27.9 \% &  42.5 \%  &  0 \%   & 30.7 \%               \\                  
                  & 100 \%&  27.1 \% &  41.5 \%  &  0 \%   & 28.7 \%               \\
16-12-6-1 &  67 \% &  24.7 \% &  41.8 \%  &  0 \%   & 28.4 \%               \\
                  &  50 \% &  28.1 \% &  42.1 \%  &  0 \%   & 30.6 \%               \\
                  &  25 \% &  31.5 \% &  44.7 \%  &  0 \%   & 35.0 \%               \\
\hline \hline                                                   
                  &Total    &  28.7 \% &  42.0 \%  &  0 \%   & 31.7 \%               \\                  
                  & 100 \%&  25.9 \% &  42.8 \%  &  0 \%   & 30.0 \%               \\
16-12-1    &  67 \% &  24.7 \% &  42.5 \%  &  0 \%   & 30.6\%               \\
                  &  50 \% &  28.1 \% &  41.5 \%  &  0 \%   & 31.8 \%               \\
                  &  25 \% &  36.2 \% &  41.2 \%  &  0 \%   & 34.3 \%               \\
\hline 
\end{tabular}
\end{table}

\begin{figure}
\centering
\includegraphics[width=0.95\columnwidth,keepaspectratio]{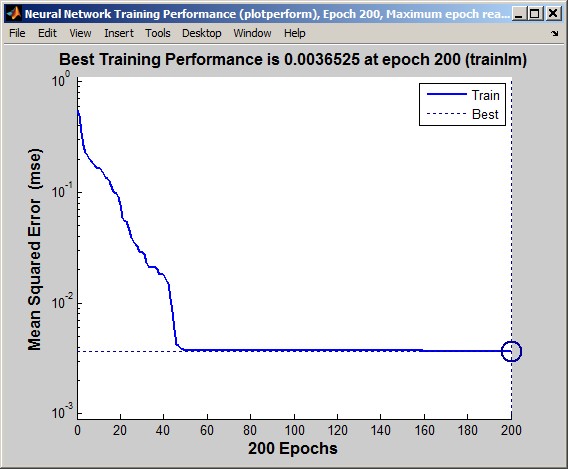}
\caption{Training performance for the 3-layer network 12-10-1.}
\label{training}
\end{figure}

\begin{table}[b]
\centering\caption{\label{nets} Error rates for 3-layer networks with trainlm algorithm}

\begin{tabular}{|c|c|c|c|c|c|}
\hline
 ANN& Total & 100 \%& 67 \% & 50 \% & 25 \%  \\
\hline \hline                                                   
 16-12-1   &  0.0 \% &   0.0 \% &  0.0 \%  &  0.0 \%   & 0.0 \%               \\               
 14-12-1   &  0.0 \% &   0.0 \% &  0.0 \%  &  0.0 \%   & 0.0 \%               \\
 14-10-1   &  0.0 \%  &  0.0 \% &  0.0 \%  &  0.0 \%   & 0.0 \%               \\
 12-10-1   &  0.0 \%  &  0.0 \% &  0.0 \%  &  0.0 \%   & 0.0 \%               \\
  12-8-1    &  1.56 \% &  1.56 \% & 1.56 \%  & 1.56 \%   & 1.56 \%               \\                                                
  10-8-1    & 2.81 \%  &  3.70 \% &  2.50 \%  &  1.25 \%   & 3.70 \%               \\                  
  10-6-1    & 1.56 \%&  1.56 \% &  1.56 \%  &  1.56 \%   & 1.56 \%               \\
   8-6-1      & 3.28 \% & 3.75 \% & 2.81 \%  &  2.81 \%   & 3.75 \%               \\
    8-4-1     & 6.71 \% &  5.31 \% & 5.61 \%  & 6.25 \%   & 9.68 \%               \\
\hline 
\end{tabular}
\end{table}

The database for training was built from real Auger ADC traces 
triggered by either Threshold trigger
(T1 - 3-fold coincidences for a single time bin for simultaneous 
signals above 1.75 VEM) or by
ToT trigger (at least 13 sub-triggers of any 2-fold coincidences 
in 120 $\mu$s interval, 
for signals 0.2 VEM above pedestal) \cite{trigger}. 
Signals detected by the T1 are relatively strong. 
The fact that the Pierre Auger
Observatory did not register up to now any event potentially 
generated by a neutrino suggests that the
thresholds for standard triggers may be too high. 
So, we need to teach the network to recognize patterns
with much lower amplitudes. We extended the database artificially 
by reducing the amplitude of real ADC traces
 (by factors 0.67, 0.5 and 0.25, respectively), 
keeping the same pedestals and shapes. 
Table \ref{nets} shows that all networks recognize traces 
with reduced amplitude pretty well.

The 12-10-1 network offers the best performance with a minimal resource occupancy (Fig. \ref{training}, however it requires 23 neurons.

Due to a limited amount of DSP blocks, we could use this network 
for a single PMT only. 
The Quartus\textsuperscript{\textregistered} compiler allows 
a compilation with arbitrarily selected implementation 
of the multipliers: either in the DSP blocks 
or in logic elements only. An implementation of the multipliers 
in the Adaptive Logic Modules (ALMs) 
is much more resource-consuming (1247 ALMs instead 
of 107 ALMs + 8 DSP blocks), however such a selection
allows an implementation of a more complicated network, 
which provides a similar performance 
(keeps approximately the same speed). 
The 3-channel 12-10-1 network needs 36 neurons 
(the 1st layer implemented in the DSP blocks) + 
33 neurons implemented in 41151 ALMs (36.5\% of 5CEFA9F31I7).

The 12-10-1 network provides also a pretty fast convergence. 
A teaching process can be accomplished in several tens of epochs.
Results from Tables \ref{methods} and \ref{nets} were obtained when networks were taught on selected patterns.
However, teaching a bigger set of real Auger data for inclined showers (and others as references) gave rather surprising 
results that the network 12-8-1 provided better 
pattern recognition than 12-10-1.

\section{FPGA implementation}

\begin{figure}[!b]
\centering
\includegraphics[width=\columnwidth,keepaspectratio]
{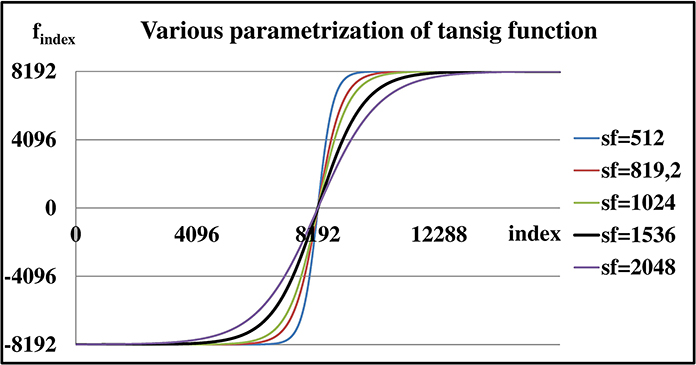}
\caption{Tested representations of the tansig function 
for the best optimization (Eq. \ref{tansig_algorithm}). }
\label{tansig}
\end{figure}

\begin{figure}[h]
\centering
\includegraphics[width=\columnwidth,keepaspectratio]{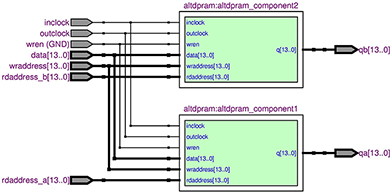}
\caption{3-port RAM as a storage bank for tansig coefficients. }
\label{ram3port}
\end{figure}

A 16-input neuron for 14-bit data and 14-bit coefficients 
is shown in Fig. \ref{neuron}.
A neuron output drives a neural transfer function - a tansig, 
which calculates a layer's output from its net input.
It can be implemented as ROM in embedded FPGA memory. 
We selected a 14-bit input, 14-bit-output tansig
implementation in RAM: a 3-port function with blocked writing 
left port to keep a reasonable
compromise between a calculation accuracy and a memory size. 
Unfortunately, ROM: a 2-port failed. The same array of 
coefficients is used for two independent neuron transfer 
functions (Fig. \ref{ram3port}).

The network was trained using 160 inclined and normal ADC traces 
(768 samples per trace). This gives 2*122880 patterns. 
For 160 inclined traces the network 12-8-1 recognized 139 
inclined showers and only 27 patterns from a reference set 
(this rate should have been zero). However, taking into account 
the number of all patterns the rate of missing traces is 0.017\% 
and the rate of wrongly recognized patterns 0.022\%. 

The fundamental algorithm for each neuron is as follows:
\begin{equation}
Neuron_{out} = \sum\limits_{k=0}^{k-1} ADC_k \cdot C_{k,layer} + bias_{k,layer}
\label{neuron_algorithm}
\end{equation}

Neuron outputs are next scaled by a transfer function which 
is chosen to have a number of properties which either enhance 
or simplify the network containing the neuron. MATLAB offers 
the hyperbolic tangent sigmoid transfer (tansig) function. 
On-line calculation of the tansig in the FPGA is not necessary, 
it is enough to store in the ROM previously
calculated values and to use the neuron output as addresses 
to the ROM.
In order to keep a sufficient accuracy with a reasonable 
size of the embedded memory we used a 16384-word dual-port 
ROM with 14-bit output. For the network 12-8-1 we had to use 
10 dual-port ROMs,
which utilized 2240 kB of embedded memory (the output from 
the last layer was given directly to a comparator).
Various parameterizations (Fig. \ref{tansig}) were tested 
for the best optimization.
The best variant for the data used was with the scaling 
factor sf = 1536 which corresponds to the range of (-5.33,...+5.33)
of the tansig argument (Fig. \ref{tansig}):
\begin{equation}
f_{index} = \frac{2}{1+e^{-2\cdot \frac{index-8192}{sf}}} - 1.
\label{tansig_algorithm}
\end{equation}

ADC samples drive the 12-bit shift register whose output 
of sequential registers are connected to neuron inputs 
(Fig. \ref{neuron}).
MATLAB provides a set of floating point coefficients 
obtained after a teaching process. 
In our practical implementation the FPGA uses the fixed 
point representation (FPR) to provide a fast enough 
registered performance and to utilize a reasonable amount 
of resources. There is no need to use floating point 
representation although Altera provides appropriate 
library procedures. For 12-bit input data at least 
2 embedded DSP multipliers have to be used for a single 
multiplication 
in Eq. \ref{neuron_algorithm}. The maximal width of the 
coefficients is 20-bit. However, we selected 18-bit 
coefficients to obtain
a 32-bit width of neuron output (Fig. \ref{neuron}).

\begin{table}[h]
\centering\caption{\label{factors}Scaling, suppression and shift factors}
\begin{tabular}{|c||c|c|c|c|c|c|c|}
\hline
 Layer& SFS & SFL &  SFX & SFB  & SHP & SHN  \\
\hline \hline                                                   
    1   &  2 &   131 072 &  8  & 524 288 & -    & 6               \\       \hline        
    2   &  4 &    32 768  &  8  & 32 768   & 14 & 1               \\      \hline         
    3   &  2 &    32 768  &  2  & 32 768   & 13  & 1               \\    
\hline 
\end{tabular}
\end{table}

 All coefficients given by MATLAB have to be converted from 
a floating-point to fixed-point 
representation in two-component code.  A simple conversion 
into two-component code is a multiplication
of data by a fixed-point scaling factor (FPSF) and an addition 
of 2*FPSF for negative values. A condition 
is that the data must be in the (-1.0,...+1.0) range. 
Table \ref{factors} shows all factors for scaling, 
suppressions and finally shifts of data. 
At first, coefficients (coeff and bias) calculated by 
MATLAB are suppressed (by factors SFS and SFX, 
respectively, to get a range (-1.0,...+1.0)
(Eq. \ref{coeff}). Next, they are scaled by factors 
SFL and SFB, respectively (Eq. \ref{bias}):
\begin{equation}
coeff_{k,layer,fixed-point} = \frac{coeff_{k,layer}}{SFS} *SFL
\label{coeff}
\end{equation}
\begin{equation}
bias_{k,layer,fixed-point} = \frac{bias_{k,layer}}{SFX} *SFB.
\label{bias}
\end{equation}

The 32-bit signed output of the neuron (starting from the 2nd layer) 
is shifted right before a summation with bias due to
very high values from the tansig transfer function 
(mostly either $\sim$-8192 or $\sim$8191):
\begin{equation}
P =\left ( \sum\limits_{k=0}^{k-1} ADC_k \cdot C_{k,layer} \right ) >> SHP.
\label{shift}
\end{equation}

Addresses for the tansig function are additionally optimized 
to use the most sensitive function response region:
\begin{equation}
address_{k,layer} = (P >> SHN) + 8192.
\label{address}
\end{equation}

The highest bits from the neuron  (Eq. \ref{neuron_algorithm}) 
are neglected as irrelevant for a big argument of the tansig 
transfer function. Addresses are cropped to the range 
$< 0,...,16383 >$.

In order to save some amount of memory we used
 Altera\textsuperscript{\textregistered} library RAM: 
3-port (as dual output ROM) with an initiation file, 
and blocked writing to the left port. The library function ROM: 
2-port unfortunately failed.
The same coefficient array is used for two independent addresses 
and gives two independent tansig
coefficients. The RAM: 3-port saves twice embedded M9K memory 
blocks in comparison to a
simple implementation of the ROM: 1-port library function.

\begin{figure}[h]
\centering
\includegraphics[width=\columnwidth,keepaspectratio]{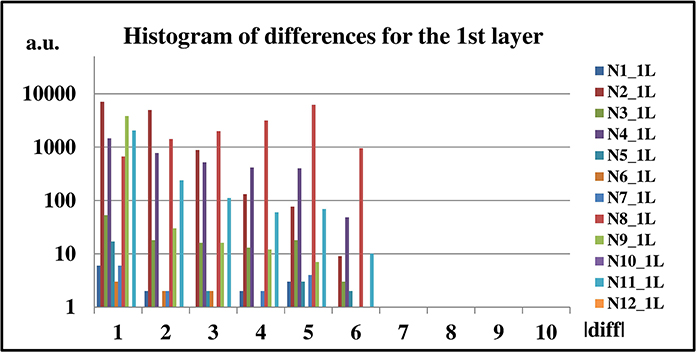}
\includegraphics[width=\columnwidth,keepaspectratio]{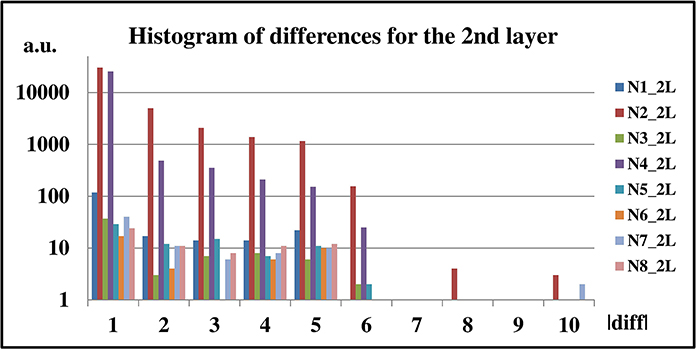}
\caption{Histogram of differences between the tansig integer output (from RAM: 3-PORT)
and exact calculation for the 1st and 2nd layers. }
\label{diff}
\end{figure}

An analysis of differences between the output data from 
neurons shows that differences reach
a maximal value of only 1 ADC-unit. However, due to the 
relatively sharp slope of the tansig function in the 
central range, an error of 1 ADC-unit generates an output 
error of up to 6 ADC-units for the next (2nd) layer
and even 10 ADC-units from the 2nd to the 3rd layer (Fig. \ref{diff}).
Nevertheless, the final error is negligible. A comparison 
of registered patterns for inclined showers (161/160) 
or spuriously recognized patterns for reference traces (39/160) 
shows that they are exactly the same as for the exact calculation
(with double precision representation) and for the FPGA 
calculation in fixed-point representation with 
optimized bus and coefficients widths.

\begin{figure}[h]
\centering
\includegraphics[width=\columnwidth,keepaspectratio]{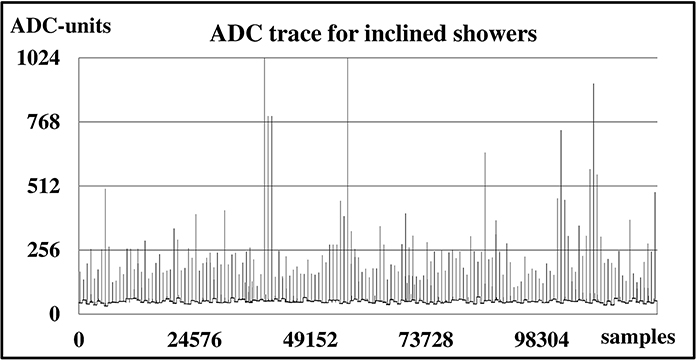}
\includegraphics[width=\columnwidth,keepaspectratio]{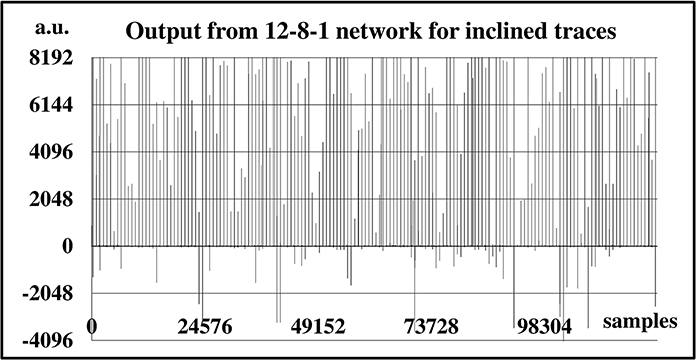}
\caption{ADC trace for positive-marked inclined showers (from 160 events) (upper graph) 
and corresponding output for 12-8-1 neural network (lower graph). }
\label{incl}
\end{figure}

\begin{figure}[h]
\centering
\includegraphics[width=\columnwidth,keepaspectratio]{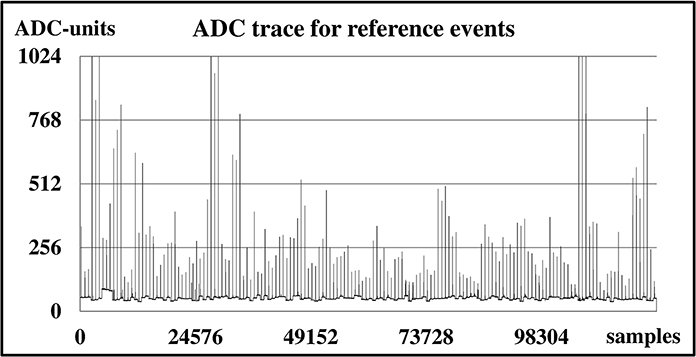}
\includegraphics[width=\columnwidth,keepaspectratio]{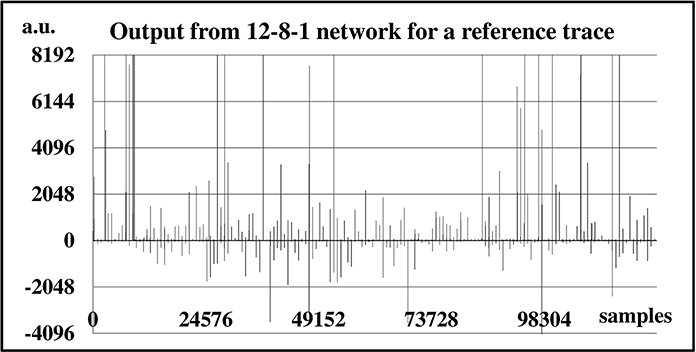}
\caption{ADC trace negative-marked reference showers 
(from 160 events) (upper graph) 
and corresponding output for 12-8-1 neural network. }
\label{norm}
\end{figure}

\section{Simulations}

The structure of the neuron network has been implemented 
in several FPGA families: 
Cyclone\textsuperscript{\textregistered} III, 
Stratix\textsuperscript{\textregistered} III and
Cyclone\textsuperscript{\textregistered} V. 
The response of neural network on trained patterns 
was verified for 16-point inputs (taken from shift registers,
where ADC data were permanently delayed sequentially) 
with fixed coefficients calculated in MATLAB.
For the biggest FPGA from 
Cyclone\textsuperscript{\textregistered} III family 
(EP3C120F780C7) 
the multipliers in the neuron from the last layer 
have to be implemented in the logic cells due to a lack 
of DSP embedded blocks. This reduced the speed below 
our requirements. The middle-size FPGA EP3SL150F780C2 
was a perfect chip for 
Quartus\textsuperscript{\textregistered} simulation.
We decided to make simulations using a relatively old tool: 
the Quartus\textsuperscript{\textregistered} 
simulator as a much faster tool than the currently
recommended ModelSim.

Figs. \ref{incl} and \ref{norm} show results of simulations 
for inclined (positive marker) and reference (negative marker)
traces. For trained patterns the recognition is almost perfect. 
On 160 events with positive markers (totally  122 760 samples)
161 patterns were recognized by the 12-8-1 network 
(only a single false event - Fig. \ref{incl}).
On 160 reference events (with negative markers) 39 spurious 
events were registered, however, 12 of which had very high 
amplitude, which would have been also registered 
by the standard trigger.

Results of simulations confirm that the noise is perfectly 
rejected. On the output of the 3rd layer, the simple comparator
was used instead of the tansig procedure (with an embedded memory).

\section{Laboratory tests}

The surface detector electronics is being improved from 10-bit 
40 MSps to at least 12-bit 120/160 MSps ADCs. 
The University of \L{}\'od\'{z} has been developing the new 
Front-End Board based on the 
Altera\textsuperscript{\textregistered} 
Cyclone\textsuperscript{\textregistered} V 5CEFA9F31I7 and 
8 channels supported by the ADS4249  (Texas Instr. 2-channel, 
14-bits 250MSps ADCs) \cite{RT2014-ANN}.
 It can fully test the developing ANN
also under real environmental conditions in the Argentinean pampas. 

\begin{figure}[t]
\centering
\includegraphics[width=0.95\columnwidth,keepaspectratio] {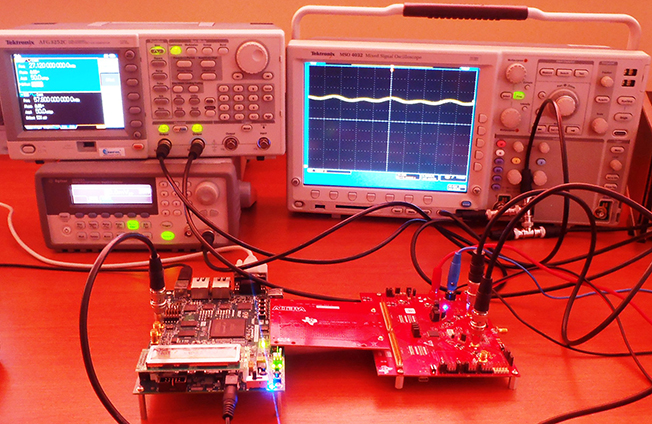}
\caption{Experimental setup with arbitrary pattern generators 
Tektronix AFG3252C and Agilent 33250A,
Altera\textsuperscript{\textregistered} 
Cyclone\textsuperscript{\textregistered} V E development kit, 
Altera\textsuperscript{\textregistered} HSMC-ADC-BRIDGE 
and Texas Instr. ADS4249EVM.}
\label{lab-setup}
\end{figure}

Before the field tests we were running laboratory tests 
based on the Altera development kit 
DK-DEV-5CEA7N driven via HSMC-ADC-BRIDGE from the ADS4249 
Evaluation Module (EVM). The ADC on the EVM
is driven from the two channel arbitrary function generator 
Tektronix AFG3252. The first channel generates patterns
corresponding to the "old" showers (marker "+"), the second 
one generates reference traces (marker "-").
Channels are uncorrelated, they run with different 
frequencies and duty cycles (Fig. \ref{lab-setup}).

The FPGA trigger (either simple T1 or DCT based) freezes 
incoming traces and sends their output
via the UART in NIOS to the PC. The virtual processor 
stores several hundred patterns in the RAM (both the developed 
FEB and the development kit
contain a large enough external SDRAM). Thus, it starts 
the learning process with the algorithm extracted from 
the MATLAB package.
Calculated coefficients are sequentially sent to the 
temporary D registers in the FPGA fast logic and are next 
simultaneously (in a single clock cycle) reloaded to the 
final registers driving the multipliers.
Trigger rates for positively vs. negatively marked patterns 
agreed with our theoretical simulations.


\section{Conclusion}

A huge amount of progress in electronics allows an introduction 
of new, much more powerful FPGAs and
an implementation of much more sophisticated mathematical 
algorithms for real time processes.
Neutrino physics is one of the discipline where new developments, 
in both hardware, and software, can significantly improve
the efficiency of rare event detection. It is especially 
interesting where theories estimate neutrino fluxes over 
a very wide range.

The spectral trigger based on the Discrete Cosine Transform, 
offering a pattern recognition technique, has been already
implemented in a test Front-End Board and tested in a test 
detector in Malarg\"ue (Argentina). The new electronics 
developed according to the Auger-Beyond-2015 task in 
the Pierre Auger upgrade project allows an implementation 
of both DCT and ANN algorithms. 

The Technical Board of the Pierre Auger Collaboration
selected 8 surface detectors (a hexagon + twin in the center 
for an investigation of possible GPS jitter)
in a north-west region of the SD array  for tests of the 
new FEB on the Cyclone\textsuperscript{\textregistered} V platform 
with 3-4 times higher sampling and 14-bit resolution with 
a cooperation of the SPMT and possibly other detectors. 
Simultaneously, the DCT triggers will be implemented in parallel 
with the standard ones to verify the detection of 
very inclined showers based on an on-line analysis 
of a shape of signals in the frequency domain.

This platform is appropriate also for tests of artificial 
neural networks. The biggest FPGA chip 5CEFA9F31I7N 
in the new Front-End allows an implementation of two 
12-8-1 networks with multipliers fully embedded in DSP blocks.
Three PMTs require 3 networks. The FPGA is big enough 
to implement mixed DSP/fast logic multipliers for 3 networks. 

We run CORSIKA \cite{CORSIKA} simulation for proton, 
iron, $\nu_{\mu}$ and $\nu_{\tau}$ primaries. Output data 
collected at 1450 m (the level of the Pierre Auger Observatory) 
was an input for OffLine \cite{OffLine} providing the ADC response 
in the WCDs.
The obtained ADC traces were thus used to train the 12-8-1 
neural network implemented already in the 5CEFA7F31I7
FPGA on the Cyclone\textsuperscript{\textregistered} V 
development kit. This FPGA is a smaller 
version of the chip being designed for the Front-End Board 
for the Auger-Beyond-2015 task.
Preliminary results show that the 16-point ANN algorithm 
can detect neutrino events currently neglected 
by the standard Auger
triggers and can support a recognition of neutrino-induced 
very inclined showers 
when ADC traces are relatively short and the muonic bump 
is better separated from the electromagnetic component.


\section*{Acknowledgment}


The authors would like to thank the Pierre Auger Collaboration for an opportunity
of using the CORSIKA and OffLine simulation packages.

\ifCLASSOPTIONcaptionsoff
  \newpage
\fi


\begin{thebibliography}{1}

\bibitem{Nagano}  M. Nagano and A. A. Watson, "Observations and implications
of the ultrahigh-energy cosmic rays", \emph{Rev. of Modern Phys.}, vol. 72, no. 3, pp. 689-732, 2000.

\bibitem{Halzen} F. Halzen and D. Hooper, "High-energy neutrino astronomy:
the cosmic ray connection", \emph{Rep. on Progress in Phys.}, vol. 65, no. 7, p. 1025, 2002.

\bibitem{Becker}  J. K. Becker, "High-energy neutrinos in the context of multi-messenger
astrophysics", \emph{Phys. Reports}, vol. 458, no. 4-5, pp. 173–246, 2008.

\bibitem{Sigl} P. Bhattacharjee and G. Sigl, "Origin and propagation of
extremely high-energy cosmic rays", \emph{Phys. Reports}, vol. 327, no. 3-4, pp. 109-247, 2000.

\bibitem{photon_limit} J. Abraham et al. [Pierre Auger Collaboration], 
''Upper limit on the cosmic-ray photon fraction at EeV energies from the Pierre Auger Observatory'',
\emph{Astropart. Phys.}, vol. 31, pp. 399-406, 2009.

\bibitem{GZK} K. Greisen, "End to the cosmic-ray spectrum?", \emph{Phys. Rev. Lett.}, vol. 16, pp. 748-750, 1966.

\bibitem{ZK} 
G. T. Zatsepin and V. A. Kuzmin, "Upper limit of the spectrum
of cosmic rays",  \emph{Pis’ma v Zhurnal Eksperimentalnoi i Teoreticheskoi Fiziki}, vol. 4, p. 114, 1966, 
English translation in: \emph{JETP Letters}, vol. 4, p. 78, 1966.

\bibitem{Abbasi} [Hi-Res Fly's Eye Collaboration], "First Observation of the Greisen-Zatsepin-Kuzmin Suppression",
\emph{Phys. Rev. Lett.}, vol. 100, no. 10, article 101101, 5 pages, 2008.

\bibitem{PAO} [Pierre Auger Collaboration], "Properties and performance of the prototype instrument 
for the Pierre Auger Observatory", \emph{Nucl. Instr. and Meth.} ser. A, vol. 523, no. 1-2, pp. 50–95, 2004.

\bibitem{PAO-GZK} 
[Pierre Auger Collaboration], “Observation of the suppression of the flux of cosmic rays above 
$4\cdot10^{19}$ eV,” Physical Review Letters, vol. 101, no. 6, article 061101, 7 pages, 2008.

\bibitem{Olinto} K. Kotera, D. Allard, and A. V. Olinto, "Cosmogenic neutrinos: parameter space and detectability from PeV to ZeV",
\emph{JCAP}, vol. 2010,no. 10, article 013, 2010.

\bibitem{Seckel} D. Seckel and T. Stanev, "Neutrinos: the key to ultrahigh energy cosmic rays",
\emph{Phys. Rev. Lett.}, vol. 95, no. 14, article 141101, 3 pages, 2005.

\bibitem{Zas} E. Zas, "Neutrino detection with inclined air showers", \emph{New Journal of Phys.}, vol. 7, p. 130, 2005.

\bibitem{Smirnov} V. S. Berezinsky, A.Yu. Smirnov, "Cosmic neutrinos of ultrahigh energies and detection possibility",
\emph{Astrophys. and Space Sci.}, vol. 32, no. 2, pp. 461–482, 1975.

\bibitem{AHEP} [Pierre Auger Collaboration], 
"Ultrahigh Energy Neutrinos at the Pierre Auger Observatory",
\emph{Adv. in High Energy Phys.} vol. 2013, Article ID 708680, 18 pages.

\bibitem{Capelle} K. S. Capelle, J. W. Cronin, G. Parente, and E. Zas, 
"On the detection of ultra high energy neutrinos with the Auger Observatory",  
\emph{Astropart. Phys.}, vol. 8, no. 4, pp. 321–328, 1998.

\bibitem{Bertou} X. Bertou, P. Billoir, O. Deligny, C. Lachaud, and A. Letessier-Selvon, 
"Tau neutrinos in the Auger Observatory:  a new window to UHECR sources",
\emph{Astropart. Phys.}, vol. 17, no. 2, pp. 183–193, 2002.

\bibitem{tau} [Pierre Auger Collaboration], 
"Upper limit on the diffuse flux of ultrahigh energy tau neutrinos  from the Pierre Auger Observatory",
\emph{Phys. Rev. Lett.}, vol. 100, no. 21, article 211101, 2008.

\bibitem{DCT-NIM} Z. Szadkowski, 
``A spectral 1$^{st}$ level FPGA trigger for detection of very inclined showers based on a 16-point Discrete Cosine Transform 
for the Pierre Auger Observatory'', \emph{Nucl. Instr. Meth.}, ser. A, vol. {\bf 606}, pp. 330-343, July 2009.

\bibitem{CycloneIII}Z. Szadkowski, "Trigger Board for the Auger Surface Detector with 100 MHz
  Sampling and Discrete Cosine Transform'', \emph{IEEE Trans. on Nucl. Science}, vol. 58, pp. 1692-1700, Aug. 2011.

\bibitem{TNS2013-DCT}
Z. Szadkowski, "Optimization of the Detection of Very Inclined Showers Using a Spectral DCT Trigger
in Arrays of Surface Detectors", \emph{IEEE Trans. on Nucl. Science}, vol. 60, no. 5, pp. 3647-3653, Oct. 2013.

\bibitem{CORSIKA} https://web.ikp.kit.edu/corsika/

\bibitem{OffLine} S. Argiro et al., The offline software framework of the Pierre Auger Observatory,
\emph{Nucl. Instrum. and Meth.} ser. A, vol. 580, Issue. 3, pp. 1485-1496, Oct. 2007.

\bibitem{trigger} [Pierre Auger Collaboration],
"Trigger and aperture of the surface detector array of the Pierre Auger Observatory", 
\emph{Nucl. Instr. Meth.}, ser. A, vol. 613, pp. 29-39, Aug. 2010.

\bibitem{RT2014-ANN} Z. Szadkowski,
Front-End Board with Cyclone\textsuperscript{\textregistered}  V 
as a Test High-Resolution Platform for the 
Auger\_Beyond\_2015 Front End Electronics,
Contribution to the \emph{IEEE Real Time Conference}, 
Nara (Japan), May 26-30, 2014.

\end{thebibliography}
\end{document}